\documentstyle[twocolumn,prb,aps,epsfig,amsfonts]{revtex}

\begin{document}
\draft
\preprint{}

\newcommand{\1}{{\bf \scriptstyle 1}\!\!{1}}
\newcommand{\I}{{\rm i}}
\newcommand{\p}{\partial}
\newcommand{\D}{^{\dagger}}
\newcommand{\bx}{{\bf x}}
\newcommand{\bk}{{\bf k}}
\newcommand{\bv}{{\bf v}}
\newcommand{\bp}{{\bf p}}
\newcommand{\bu}{{\bf u}}
\newcommand{\bA}{{\bf A}}
\newcommand{\bB}{{\bf B}}
\newcommand{\bF}{{\bf F}}
\newcommand{\bI}{{\bf I}}
\newcommand{\bK}{{\bf K}}
\newcommand{\bL}{{\bf L}}
\newcommand{\bP}{{\bf P}}
\newcommand{\bQ}{{\bf Q}}
\newcommand{\bS}{{\bf S}}
\newcommand{\bH}{{\bf H}}
\newcommand{\balpha}{\mbox{\boldmath $\alpha$}}
\newcommand{\bsigma}{\mbox{\boldmath $\sigma$}}
\newcommand{\bSigma}{\mbox{\boldmath $\Sigma$}}
\newcommand{\bOmega}{\mbox{\boldmath $\Omega$}}
\newcommand{\bpi}{\mbox{\boldmath $\pi$}}
\newcommand{\bphi}{\mbox{\boldmath $\phi$}}
\newcommand{\bnabla}{\mbox{\boldmath $\nabla$}}
\newcommand{\bmu}{\mbox{\boldmath $\mu$}}
\newcommand{\bepsilon}{\mbox{\boldmath $\epsilon$}}

\newcommand{\iLambda}{{\it \Lambda}}
\newcommand{\cA}{{\cal A}}
\newcommand{\cD}{{\cal D}}
\newcommand{\cF}{{\cal F}}
\newcommand{\cL}{{\cal L}}
\newcommand{\cH}{{\cal H}}
\newcommand{\cI}{{\cal I}}
\newcommand{\cM}{{\cal M}}
\newcommand{\cO}{{\cal O}}
\newcommand{\cR}{{\cal R}}
\newcommand{\cU}{{\cal U}}
\newcommand{\cT}{{\cal T}}

\newcommand{\be}{\begin{equation}}
\newcommand{\ee}{\end{equation}}
\newcommand{\bea}{\begin{eqnarray}}
\newcommand{\eea}{\end{eqnarray}}
\newcommand{\beqa}{\begin{eqnarray*}}
\newcommand{\eeqa}{\end{eqnarray*}}
\newcommand{\nn}{\nonumber}
\newcommand{\DD}{\displaystyle}

\newcommand{\ba}{\left[\begin{array}{c}}
\newcommand{\baa}{\left[\begin{array}{cc}}
\newcommand{\baaa}{\left[\begin{array}{ccc}}
\newcommand{\baaaa}{\left[\begin{array}{cccc}}
\newcommand{\ea}{\end{array}\right]}

\twocolumn[
\hsize\textwidth\columnwidth\hsize\csname
@twocolumnfalse\endcsname

\title{The Grover algorithm with large nuclear spins in
semiconductors}

\author{Michael N.~Leuenberger
}
\address{Department of Physics and Astronomy, University of Iowa \\
IATL, Iowa, IA 52242, USA}
\author{Daniel Loss
}
\address{Department of Physics and Astronomy, University of Basel \\ 
Klingelbergstrasse 82, 4056 Basel, Switzerland}

\date{\today}
\maketitle

\begin{abstract}
We show a possible way to implement the Grover algorithm in large
nuclear spins $1/2<I\le 9/2$ in semiconductors. The Grover sequence
is performed by means of multiphoton transitions
that distribute the spin amplitude between the nuclear spin states.
They are distinguishable due to the quadrupolar splitting,
which makes the nuclear spin levels non-equidistant.
We introduce a generalized rotating frame for 
an effective Hamiltonian that governs the non-perturbative
time evolution of the nuclear spin states for arbitrary spin
lengths $I$. 
The larger the quadrupolar splitting, the better the agreement
between our approximative method using the generalized rotating
frame and exact numerical calculations.
\end{abstract}

]
\narrowtext

%start PRB

\section{Introduction}

Recent experiments\cite{Kikkawa1998,Kikkawa1999} have shown that electron spins in GaAs can preserve their coherence for distances of more than 100 $\mu$m
and for times of up to 130 ns.
Long coherence lengths and times are the main requirement for performing logic operations
in spintronics\cite{Prinz,Wolf,Awschalom,Ziese,Chtchelkanova}
and quantum computing,\cite{Awschalom,Loss,Nielsen,Bouwmeester,Braunstein,Williams,Berman,Lo,Kitaev,Johnson}
because both research fields are interested in the complete control over the phase information
of the spins and qubits, respectively.
Among the various available information carriers, nuclear spins have one of the longest coherence times
due to their weak interaction with the environment.
which amounts to storing phase information for a long time in e.g. semiconductor structures.
Either they can be used as qubits or in the unary representation.\cite{Leuenberger2002}
Although implementations based on the latter are not scalable,
they are more feasible, more robust, and can also
be implemented in classical systems.\cite{Leuenberger2003}
Coherent access to the nuclear spins in semiconductors is achieved by the all-optical NMR method
shown in Refs.~\onlinecite{Kikkawa2000} and \onlinecite{Salis}, which makes use of the hyperfine interaction
between the electron and nuclear spins and relies on the large coherence time of the electrons.
A further possibility to access the nuclear spins coherently is conventional NMR
using coils.\cite{Abragam} However, compared to the all-optical method, 
coils do not provide the spatially selective manipulation of nuclear spins.
Recently incoherent transfer of electronic spin to nuclear spin has been demonstrated experimentally
in semiconductor structures using the quantum Hall effect or ferromagnetic imprinting.\cite{Smet,Kawakami}

One of the most interesting quantum algorithms was introduced by Grover,\cite{Grover}
who demonstrated that the parallelism of unitary operations can speed up
the search for a desired quantum state.
Since the Grover algorithm needs only the superposition principle of quantum or classical mechanics,
the unary representation of an ensemble of single particles can be used,
such as the atomic levels of a beam of atoms\cite{Ahn} or the large spin of molecular magnets embedded in a crystal.\cite{Leuenberger&Loss}
In this paper we show that the NMR method presented in Ref.~\onlinecite{Leuenberger2002}
works for arbitrary spin lengths $I$. We compute the Grover algorithm 
on the single-spin states $\left|m\right>$ of the large spin of nuclear spins in semiconductors.
For this it is essential that a delocalized state $\left|\psi\right>=\sum_{m=-I}^Ia_m\left|m\right>$
with arbitrary amplitudes $a_m$ can be produced,
which we show to be feasible with a single magnetic rf pulse.
Since arbitrary amplitudes $a_m$ are needed, we derive a non-perturbative method to calculate the time evolution
of the nuclear spins that is valid for arbitrary spin lengths $I$.
We also perform exact numerical calculations based on solving the Schr\"odinger equation.
It turns out that the larger the quadrupolar splitting, the better the agreement
between our non-perturbative method and the exact numerical calculations.
In order to be able to control the nuclear spins coherently,
the energy levels $\varepsilon_m$ have to be non-equidistant,
which is ensured by the quadrupolar splitting.
Instead of encoding the information into the phases of $a_m$, we encode the information into the eigenenergies 
$\delta_{I-m}$ of the eigenstates $\left|m\right>$
in the generalized rotating frame (see below).\cite{Farhi,Grover&Sengupta}
After choosing a specific basis state $\left|M\right>$
to be looked for,
we make it degenerate with the completely delocalized state $\left|s\right>=\sum_{m=-I}^Ia_m\left|m\right>$ 
that has equal amplitudes $a_m$ in the generalized rotating frame, i.e.
a second magnetic rf pulse lets $\left|s\right>$ evolve into $\left|M\right>$, which is ensured
by the finite overlap of $\left|s\right>$ and $\left|M\right>$.

In Sec.~\ref{sec_Hamiltonian} we give the Hamiltonian that is the most suitable
for nuclear spins in semiconductors.
For the time evolution of the nuclear spins we first identify the transition (Feynman) diagrams
that give the largest contribution to the multiphoton transition amplitudes
by means of high-order time-dependent perturbation theory,
which is demonstrated in Sec.~\ref{sec_perturbative}.
From this we derive effective Hamiltonians in the so-called generalized rotating frame
that describe the time evolution of the nuclear spins
non-perturbatively for arbitrary spin lengths $I$, which is shown in Sec.~\ref{sec_generalized}.
In order to understand better our concept of the generalized rotating frame in Sec.~\ref{sec_generalized},
we first review the transformation to the standard rotating frame in Sec.~\ref{sec_standard}.
In Sec.~\ref{sec_Grover} we explain our proposed method to implement
the Grover algorithm into the single-spin system of nuclei in semiconductors.
Finally, Sec.~\ref{sec_readout} shows that conventional incoherent NMR can be used to read out
the searched state
after performing the Grover quantum search algorithm.

\begin{figure}[htb]
  \begin{center}
    \leavevmode
\epsfxsize=6cm
\epsffile{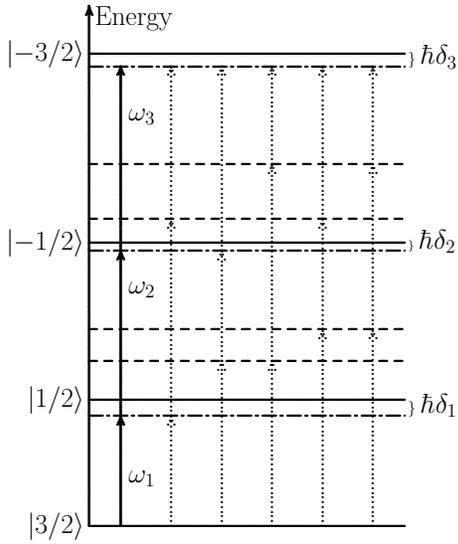}
  \end{center}
\caption{Multiphoton transition scheme (QC scheme) for the coherent population of
the $I_z$ eigenstates $\left|m\right>$ of a nuclear spin $I=3/2$.
In this scheme the Grover algorithm can be implemented non-perturbatively. 
The frequencies $\omega_k$ of the  fields
$H_k$ are red (- $\cdot$) and blue
(- -) detuned. Diagrams containing blue detunings are negligible for
large quadrupolar splitting, i.e. $A\gg\hbar\delta_k\ge 0$. 
In Sec.~\protect\ref{sec_perturbative} we explain why only the leftmost diagram is relevant.
}
\label{nuclear_spin_scheme}
\end{figure}

\section{Model Hamiltonian}
\label{sec_Hamiltonian}

The ensemble of nuclear spins in semiconductors is best described by the 
single-spin Hamiltonian
\be
\cH_0=\cH_{\rm Z}+\cH_{\rm Q},
\label{Hamiltonian_0}
\ee 
which consists of the nuclear Zeeman term
\be
\cH_{\rm Z}=-g_N\mu_NH_zI_z,
\label{Zeeman}
\ee
with the nuclear $g$-factor
$g_N=1.3$ (see Ref.~\onlinecite{Kikkawa2000}), and the quadrupolar term\cite{Abragam}
\be 
\cH_{\rm Q}=A[3I_z^2-I(I+1)].
\label{quadrupolar}
\ee 
The quadrupolar constant $A$
differs significantly between the various nuclei. For example
the all-optical NMR method shown in Ref.~\onlinecite{Salis} yields
the following quadrupolar constants for Ga and As nuclei
in GaAs semiconductors: 
$A=7\times 10^{-7}$ K for $^{69}$Ga, $A=3\times 10^{-7}$ K for $^{71}$Ga, 
and $A=2\times 10^{-6}$ K for $^{75}$As.
Let $\left|m\right>$ be the eigenstates of $\cH_0$
with eigenenergies $\varepsilon_m$.

\begin{figure}[htb]
  \begin{center}
    \leavevmode
\epsfxsize=8cm
\epsffile{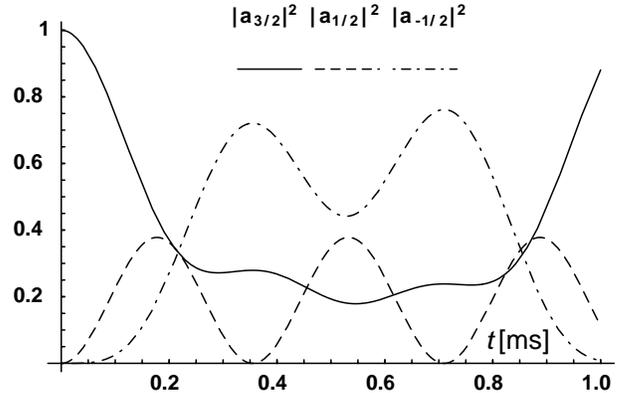}
  \end{center}
\caption{Preparation of $\left|s\right>=(1/\protect\sqrt{3})\sum_{m=-1/2}^{3/2}\left|m\right>$
by means of Eq.~(\protect\ref{general_rotating_frame_approx}) in the QC scheme, which takes about 0.2 ms for $H_1=H_2=1$ G,
$H_3=0$, $\delta_1=6083$ s$^{-1}$, and $\delta_2=0$. The analytical result is confirmed by 
numerics in Fig.~\protect\ref{QC_num}.
The calculation was done in the generalized rotating frame
as described in Sec.~\protect\ref{sec_generalized}.}
\label{QC}
\end{figure}

Next, we apply external magnetic fields $H_{x,k}(t)=\tilde{H}_k(t)\cos(\omega_kt+\Phi_k)$, $k=1,2,3,\ldots$,
that differ by the phases $\Phi_k$ and oscillate at frequencies $\omega_k$ being detuned by $\delta_k$ from
the differences between the energy levels $\varepsilon_m$.
To be more precise, $\omega_k$ is detuned by $\delta_k$ from $\varepsilon_{2I-k+1}-\varepsilon_{2I-k}$.
Fig.~\ref{nuclear_spin_scheme} shows the example for $I=3/2$ and $k=1,2,3$.
If $\delta_k<0$ the detuning is said to be blue, and if $\delta_k>0$
the detuning is said to be red.  
For  GaAs, $\omega_k,\omega_k'\sim 10$ MHz with
$\delta_k\sim 1$ kHz,
and a longitudinal magnetic field $H_z\sim 1$ T is appropriate.
It is desirable to make $H_z$ sufficiently large to accommodate many spin precessions
before the spins dephase.
The complete Hamiltonian has then the form
\be
\cH=\cH_0+V(t),
\ee
where
\be
V(t)=\sum_{k=1}^{2I} g_N\mu_N\tilde{H}_k(t)\cos(\omega_kt+\Phi_k)I_x
\ee
is the driving Hamiltonian given by the external magnetic fields.
As usual, the spin operator can be decomposed into ladder operators, i.e. $I_x=(I_++I_-)/2$.

For the implementation of the non-perturbative version of the Grover algorithm\cite{Farhi,Grover&Sengupta}
we have to be able to produce a completely delocalized state of 
the form $\left|\psi\right>=\sum_{m=-I}^Ia_m\left|m\right>$
with arbitrary amplitudes $a_m$.
Furthermore, we need to have coherent control over
the unitary time evolution of $\left|\psi(t)\right>$
for arbitrary times $t$.
We will show in the next sections that
our non-perturbative method gains complete control
over all the amplitudes $a_m(t)$,
which can be used in a future experimental implementation.

\begin{figure}[htb]
  \begin{center}
    \leavevmode
\epsfxsize=8cm
\epsffile{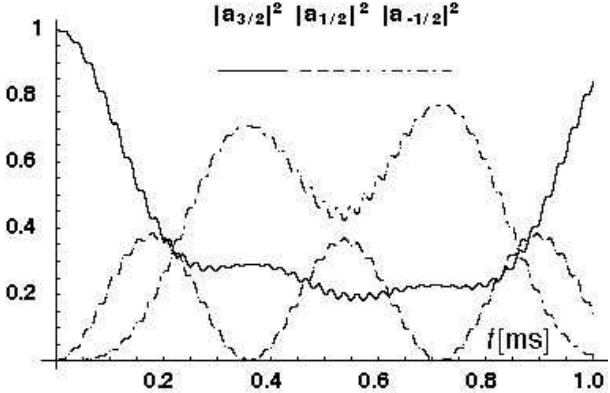}
  \end{center}
\caption{Preparation of $\left|s\right>=(1/\protect\sqrt{3})\sum_{m=-1/2}^{3/2}\left|m\right>$
by solving the Schr\"odinger equation exactly for 
$^{71}$Ga nuclei in the QC scheme, which takes about 0.2 ms for $H_1=H_2=1$ G,
$H_3=0$, $\delta_1=6083$ s$^{-1}$, and $\delta_2=0$. 
The small oscillations are due to the five diagrams with blue detunings
(see Fig.~\protect\ref{nuclear_spin_scheme})
that have been neglected in Fig.~\protect\ref{QC}.
The $^{71}$Ga nuclei have an anisotropy constant of $A=3\times 10^{-7}$ K.}
\label{QC_num}
\end{figure}

\section{Perturbative expansion}
\label{sec_perturbative}

Usually perturbation theory is useful only if the transverse term $V(t)$ is small
compared to the longitudinal term $\cH_0$. But here we use
high-order time-dependent perturbation theory
to identify the transition diagrams with the largest contribution
to the transition amplitudes 
$S_{m,m'}=\lim_{t\rightarrow -\infty,t'\rightarrow\infty}\left<m\left|U(t,t')\right|m'\right>$.
For this we first expand the $S$-matrix in powers of $V(t)$, i.e.
$S=\sum_{j=0}^\infty S^{(j)}$.
The $j$th-order term of the perturbation series of the $S$-matrix in powers of the driving Hamiltonian $V(t)$ is expressed by 
\bea
S^{(j)} & = & \left(\frac{1}{i\hbar}\right)^j\prod_{k=1}^{j-1}\int_{-\infty}^\infty
dt_k\int_{-\infty}^\infty dt_j\Theta(t_k-t_{k+1})U_0(\infty,t_1)\nn\\
& & \times V(t_1)U_0(t_1,t_2)V(t_2)\ldots V(t_j)U_0(t_j,-\infty),
\eea
which corresponds to the sum over all transition diagrams of order $j$, and where 
$U_0(t,t_0)=e^{-i\cH_0(t-t_0)/\hbar}$ is the free propagator, $\Theta(t)$ is the Heavyside function.

For illustration, we derive now a three $\sigma$-photon transition from $\left|3/2\right>$ to $\left|-3/2\right>$. 
All the relative phases $\Phi_k$ between the magnetic fields are assumed to be zero in this calculation.
Then we obtain
\bea
S_{-\frac{3}{2},\frac{3}{2}}^{(3)} & = & \left(\frac{g\mu_N}{4i\hbar}\right)^3 
\int\!\int\!\int dt_1dt_2dt_3\Theta(t_1-t_2)\Theta(t_2-t_3) \nn\\
& & \times p_{-\frac{3}{2},\frac{3}{2}}
e^{i\varepsilon_{\frac{3}{2}}t_1/\hbar}e^{-i\varepsilon_{\frac{1}{2}}(t_1-t_2)/\hbar}
e^{-i\varepsilon_{-\frac{1}{2}}(t_2-t_3)/\hbar} \nn\\
& & \times e^{-i\varepsilon_{-\frac{3}{2}}t_3/\hbar}\prod_{l=1}^3\sum_{k=1}^{3}\tilde{H}_k(t_l)\left(e^{i\omega_kt_l}+e^{-i\omega_kt_l}\right) \nn\\
& \approx & \left(\frac{g\mu_N}{4i\hbar}\right)^3 
\int\!\int\!\int dt_1dt_2dt_3\Theta(t_1-t_2)\Theta(t_2-t_3) \nn\\
& & \times p_{-\frac{3}{2},\frac{3}{2}}\tilde{H}_3(t_1)\tilde{H}_2(t_2)\tilde{H}_1(t_3)
e^{i(\omega_{-\frac{3}{2},\frac{3}{2}}-\omega_1-\omega_2-\omega_3)t_1} \nn\\
& & \times e^{-i(\omega_{-\frac{1}{2},\frac{3}{2}}-\omega_1-\omega_2)(t_1-t_2)}
e^{-i(\omega_{\frac{1}{2},\frac{3}{2}}-\omega_1)(t_2-t_3)}
\eea
where from the $3^3=27$ terms remains only one due to the rotating wave approximation and after keeping only the terms with the smallest detuning.
This diagram is the most left one in Fig.~\ref{nuclear_spin_scheme}.

\begin{figure}[htb]
  \begin{center}
    \leavevmode
\epsfxsize=8cm
\epsffile{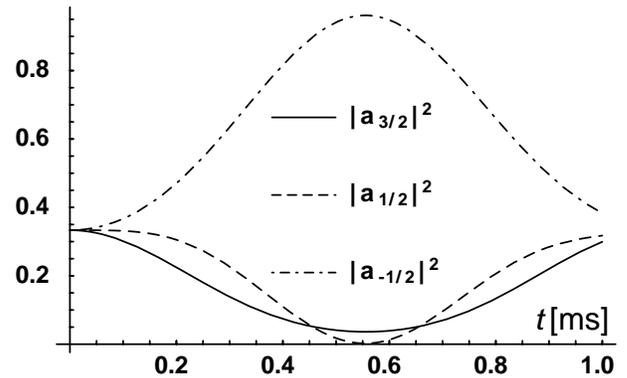}
  \end{center}
\caption{Grover algorithm calculated by means of Eq.~(\protect\ref{general_rotating_frame_approx}) 
in the QC scheme, where  
$\left|s\right>=(1/\protect\sqrt{3})\sum_{m=-1/2}^{3/2}\left|m\right>$ is concentrated mainly
into $\left|-1/2\right>$ after 0.55 ms for $H_2=\hbar\delta_2/2g_N\mu_N=1$ G, $h_1=h_2$,
$h_3=0$, $\delta_1=0$. The duration of the
QC is $\le 1/2\nu_{\rm Rabi}^{(2)}$.
This result is numerically confirmed in Fig.~\protect\ref{Grover_num}.
The calculation was done in the generalized rotating frame
as described in Sec.~\protect\ref{sec_generalized}.}
\label{Grover}
\end{figure}

Next, we make use of the relation
\bea
e^{-i(\omega_{m,m'}-\omega)(t-t')}\Theta(t-t') & = & \frac{i}{2\pi}
\int_{-\infty}^\infty d\Omega \nn\\
& & \times\frac{e^{-i\Omega(t-t')}}{\Omega-\omega_{m,m'}+\omega+i\varepsilon}
\eea
and substitute the Fourier transform
\be
\tilde{H}(\Omega)=\frac{1}{\sqrt{2\pi}}\int_{-\infty}^\infty\tilde{H}(t)e^{i\Omega t}dt,
\ee
which yields
\bea
S_{-\frac{3}{2},\frac{3}{2}}^{(3)} & \approx & \left(\frac{g\mu_N}{4\hbar}\right)^3 
\frac{p_{-\frac{3}{2},\frac{3}{2}}}{\sqrt{2\pi}i}\int\!\int d\Omega d\Omega'
\frac{\tilde{H}_2(\Omega-\Omega')\tilde{H}_3(\Omega')}
{\Omega'-\omega_{\frac{1}{2},\frac{3}{2}}+\omega_1+i\varepsilon} \nn\\
& & \times\frac{\tilde{H}_1(\omega_{-\frac{3}{2},\frac{3}{2}}-\omega_1-\omega_2-\omega_3
-\Omega)}
{\Omega-\omega_{-\frac{1}{2},\frac{3}{2}}+\omega_1+\omega_2+i\varepsilon} \nn\\
& \approx & \left(\frac{g\mu_N}{4\hbar}\right)^3 
\frac{p_{-\frac{3}{2},\frac{3}{2}}}{\sqrt{2\pi}i}\int\!\int d\Omega d\Omega'
\frac{\tilde{H}_2(\Omega-\Omega')\tilde{H}_1(\Omega')}
{\omega_{\frac{1}{2},\frac{3}{2}}+\omega_1} \nn\\
& & \times\frac{\tilde{H}_3(\omega_{-\frac{3}{2},\frac{3}{2}}-\omega_1-\omega_2-\omega_3
-\Omega)}
{\omega_{-\frac{1}{2},\frac{3}{2}}+\omega_1+\omega_2},
\eea
as long as all $\tilde{H}_k(\Omega)$ have a sufficiently narrow peak around $\Omega=0$ and $\omega_k$ are detuned. 

\begin{figure}[htb]
  \begin{center}
    \leavevmode
\epsfxsize=8cm
\epsffile{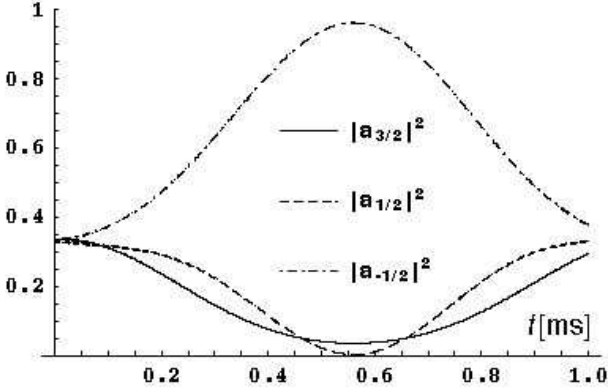}
  \end{center}
\caption{Grover algorithm calculated by solving the Schr\"odinger equation exactly
for $^{75}$As nuclei in the QC scheme, where  
$\left|s\right>=(1/\protect\sqrt{3})\sum_{m=-1/2}^{3/2}\left|m\right>$ is concentrated mainly
into $\left|-1/2\right>$ after 0.55 ms for $H_2=\hbar\delta_2/2g_N\mu_N=1$ G, $h_1=h_2$,
$h_3=0$, $\delta_1=0$. 
The $^{75}$As nuclei have an anisotropy constant of $A=2\times 10^{-6}$ K,
which is about 10 times larger than for $^{71}$Ga nuclei.
Therefore the small oscillations due to the neglected five
diagrams in Fig.~\protect\ref{nuclear_spin_scheme} are here almost invisible.
Hence we can infer that the larger the quadropular constant $A$, the better
the agreement between our non-perturbative method and exact numerics.}
\label{Grover_num}
\end{figure}

After evaluation of the three-fold convolution we obtain
\bea
S_{-\frac{3}{2},\frac{3}{2}}^{(3)} & \approx & \left(\frac{g\mu_N}{4\hbar}\right)^3 
\frac{\int dt\prod_{k=1}^3\tilde{H}_k(t)
e^{i(\omega_{-\frac{3}{2},\frac{3}{2}}-\omega_1-\omega_2-\omega_3)t}}
{ip_{-\frac{3}{2},\frac{3}{2}}^{-1}(\omega_1-\omega_{\frac{1}{2},\frac{3}{2}})(\omega_1+\omega_2-\omega_{-\frac{1}{2},\frac{3}{2}})} \nn\\
& = & \left(\frac{g\mu_N}{4\hbar}\right)^3 
\frac{\prod\limits_{k=1}^3H_k
\delta^{(T)}(\omega_{-\frac{3}{2},\frac{3}{2}}-\omega_1-\omega_2-\omega_3)}
{i(2\pi p_{-\frac{3}{2},\frac{3}{2}})^{-1}\delta_1\delta_2},
\eea
where $p_{m,m'}=\prod_{k=m}^{m'}\left<k\left|I_-\right|k+1\right>$
is the spin amplitude between $\left|k\right>$ and $\left|k+1\right>$
and 
$\delta^{(T)}(\omega)=\frac{1}{2\pi}\int_{-T/2}^{+T/2}e^{i\omega
t}dt=\frac{\sin(\omega T/2)}{\pi\omega}$ is the delta-function of width
$1/T$.
In addition, we have used rectangular pulse shapes of duration $T$ for all fields, i.e.
\be
\tilde{H}_k(t)=\left\{\begin{array}{l} H_k \text{ for } -T/2<t<T/2, \\ 0 \text{ otherwise}.\end{array}
\right.
\ee
The energy is conserved for $\omega T\gg 1$. Also, the duration
$T$ of the rf pulses must not exceed the dephasing time $\tau_{\phi}$ of 
the spin states.

If we keep also the further five diagrams shown in Fig.~\ref{nuclear_spin_scheme}, we obtain
\bea
\tilde{S}_{-\frac{3}{2},\frac{3}{2}}^{(3)} & = & 
\prod_{k=1}^3H_ke^{-i\Phi_k}
\left[\frac{1}{\delta_1\delta_2}-\frac{1}{\delta_1(\frac{6A}{\hbar}-\delta_1+\delta_2)}\right.
\nn\\
& &
-\left.\frac{1}{\frac{6A}{\hbar}+\delta_1-\delta_2}\left(
\frac{1}{\delta_2}
-\frac{1}{\frac{12A}{\hbar}+\delta_1}\right)\right.
\nn\\
& &
+\left.\frac{1}{\frac{12A}{\hbar}+\delta_2}\left(
\frac{1}{\frac{6A}{\hbar}-\delta_1+\delta_2}
+\frac{1}{\frac{12A}{\hbar}+\delta_1}\right)\right]
\label{S3}
\eea
for $\delta_3=0$,
\be
\tilde{S}_{-\frac{1}{2},\frac{3}{2}}^{(2)}=
\prod_{k=1}^2H_ke^{-i\Phi_k}
\left(-\frac{1}{\delta_1}+\frac{1}{\frac{6A}{\hbar}+\delta_1}\right)
\label{S2}
\ee
for $\delta_2=0$ and $H_3=0$, and 
\be
\tilde{S}_{\frac{1}{2},\frac{3}{2}}^{(1)}=H_1e^{-i\Phi_1}
\ee
for $\delta_1=0$ and $H_2=H_3=0$,
where
$S_{\frac{3}{2}-j,\frac{3}{2}}^{(j)}=\frac{2\pi}{i}
\left(\frac{g_N\mu_N}{4\hbar}\right)^j\tilde{S}_{\frac{3}{2}-j,\frac{3}{2}}^{(j)}
p_{\frac{3}{2}-j,\frac{3}{2}}
\delta^{(T)}(\omega_{\frac{3}{2}-j,\frac{3}{2}}-\sum_{k=1}^j\omega_k)$. 

\begin{figure}[htb]
  \begin{center}
    \leavevmode
\epsfxsize=8cm
\epsffile{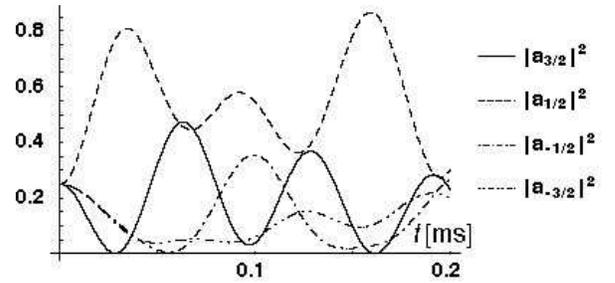}
  \end{center}
\caption{Grover algorithm calculated by means of Eq.~(\protect\ref{general_rotating_frame_approx}) 
in the QC scheme, where  
$\left|s\right>=(1/\protect\sqrt{4})\sum_{m=-3/2}^{3/2}\left|m\right>$ is concentrated mainly
into $\left|1/2\right>$ after 0.05 ms for $h_1=h_2=h_3=\hbar\delta_1/2g_N\mu_N=5$ G,
$\delta_2=\delta_3=0$. }
\label{Grover_I=32_M=12}
\end{figure}

It is interesting to note that 
\be
\lim_{A\rightarrow 0} S_{-3/2,3/2}^{(3)}=
\lim_{A\rightarrow 0} S_{-1/2,3/2}^{(2)}=0,
\ee
i.e. destructive interference is maximal. 
However, if 
\be
A\gg\hbar|\delta_k|, \quad k=1,2,3,
\label{anisotropy_condition}
\ee
destructive interference is negligible.
In other words, the most left diagram in Fig.~\ref{nuclear_spin_scheme}
gives the largest contribution to the transition amplitude.
This means that this diagram only is responsible
for the time evolution of the nuclear spins.
So Eq.~(\ref{anisotropy_condition}) is the basis condition
for finding the Hamiltonian in the generalized rotating
frame that describes the time evolution
of our nuclear spin system for arbitrary spin lengths $I$
and arbitrary times $t$ (see Sec.~\ref{sec_generalized}).

\section{The standard rotating frame}
\label{sec_standard}

Before deriving the transformation of the Hamiltonian $\cH$
including many magnetic fields $H_{x,k}(t)$ oscillating at different frequencies $\omega_k$ 
to the generalized rotating frame (see next section), 
it is instructive to have a look at the transformation
of a Hamiltonian $\cH_{\rm s}$
including only a single circularly polarized oscillating magnetic field
$\bH_\bot(t)=H_\bot(\cos\omega t{\bf e}_x+\sin\omega t{\bf e}_y)$ to the
standard rotating frame. We start from the Hamiltonian
\bea
\cH_{\rm s} & = & -g_N\mu_N[H_zI_z+H_\bot(I_x\cos\omega t+I_y\sin\omega t)] \nn\\
& = & -g_N\mu_NH_zI_z+g_N\mu_N\frac{1}{2}H_\bot e^{i\omega tI_z}I_xe^{-i\omega tI_z}.
\label{standard_Hamiltonian}
\eea
The proof for the second equality can be given as follows:
First let us define the function 
\be
f(\varphi)=e^{i\varphi I_z}I_xe^{-i\varphi I_z}.
\ee
The first and second derivative read
\bea
\frac{df}{d\varphi} & = & -e^{i\varphi I_z}I_ye^{-i\varphi I_z}, \\
\frac{d^2f}{d\varphi^2} & = & -f(\varphi).
\eea
The solution of the last differential equation is
\be
f(\varphi)=A\cos\varphi+B\sin\varphi.
\label{function}
\ee
The boundary conditions 
\be
f(0)=A=I_x,\quad \left(\frac{df}{d\varphi}\right)_{\varphi=0}=B=-I_y
\ee
reduce Eq. (\ref{function}) to
\be
f(\varphi)=I_x\cos\varphi-I_y\sin\varphi,
\ee
which proves Eq. (\ref{standard_Hamiltonian}).

\begin{figure}[htb]
  \begin{center}
    \leavevmode
\epsfxsize=8cm
\epsffile{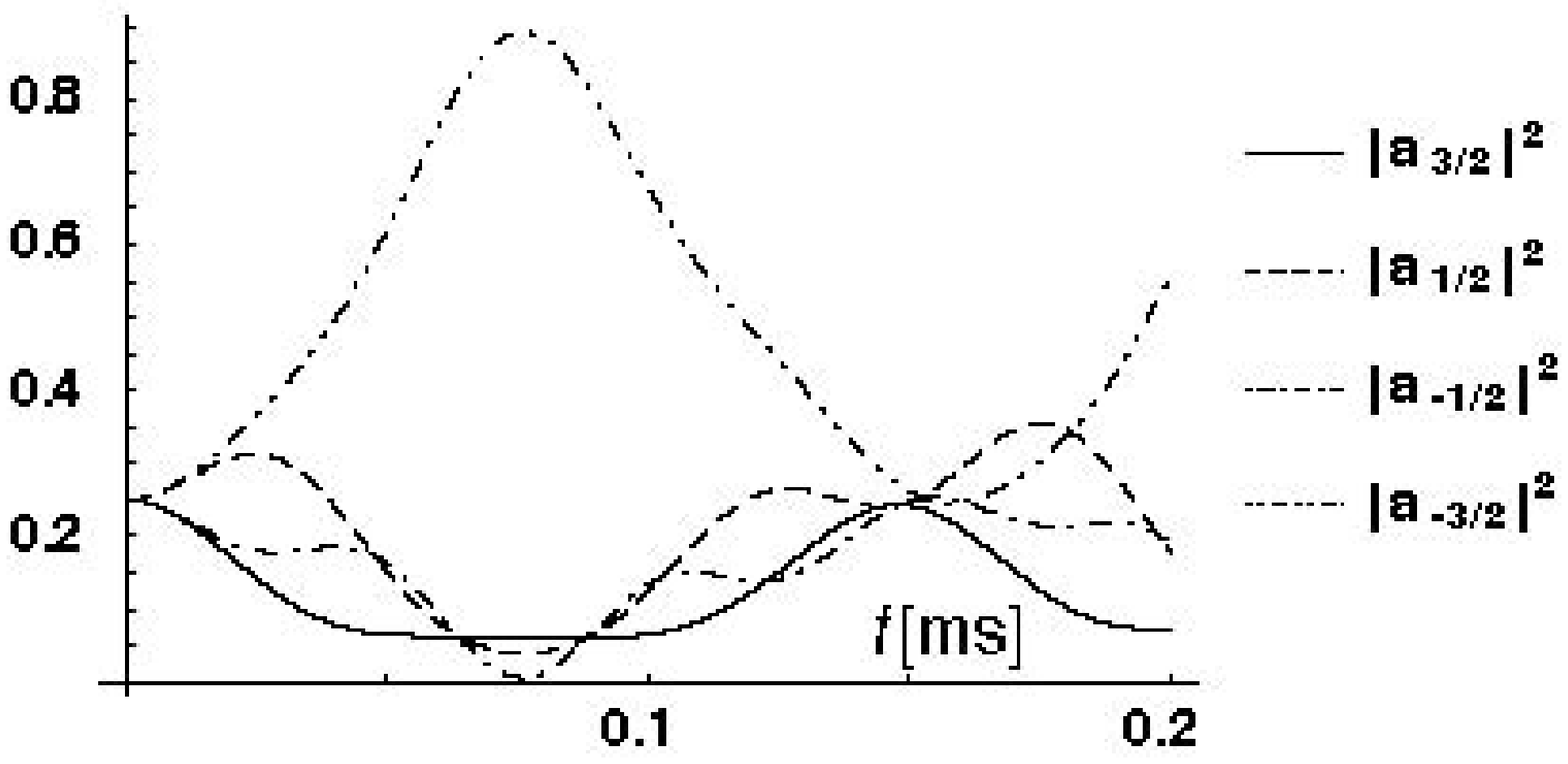}
  \end{center}
\caption{Grover algorithm calculated by means of Eq.~(\protect\ref{general_rotating_frame_approx}) 
in the QC scheme, where  
$\left|s\right>=(1/\protect\sqrt{4})\sum_{m=-3/2}^{3/2}\left|m\right>$ is concentrated mainly
into $\left|-3/2\right>$ after 0.05 ms for $h_1=h_2=h_3=\hbar\delta_3/2g_N\mu_N=5$ G,
$\delta_1=\delta_3=0$. }
\label{Grover_I=32_M=-32}
\end{figure}

In order to derive the Hamiltonian in the rotating frame, we have
to transform the Schr\"odinger equation
\be
i\hbar\frac{\p\left|\psi\right>}{\p t}=\cH_{\rm s}\left|\psi\right>
\label{Schroedinger}
\ee
with the unitary operation $U$. The left hand side of Eq. (\ref{Schroedinger}) transforms into 
\be
i\hbar\frac{\p\left|\psi\right>}{\p t} = i\hbar\left(
U^{-1}\frac{\p\left|\psi_{\rm rot}\right>}{\p t}+\frac{\p U^{-1}}{\p t}\left|\psi_{\rm rot}\right>\right),
\label{left}
\ee
and the right hand side of Eq. (\ref{Schroedinger}) into
\be
\cH\left|\psi\right>=\cH U^{-1}\left|\psi_{\rm rot}\right>,
\label{right}
\ee
where we have defined the state in the rotating frame by $\left|\psi_{\rm rot}\right>=U\left|\psi\right>$.
Combining both sides (\ref{left}) and (\ref{right}) yields
\be
i\hbar\left(
U^{-1}\frac{\p\left|\psi_{\rm rot}\right>}{\p t}+\frac{\p U^{-1}}{\p t}\left|\psi_{\rm rot}\right>\right)
=\cH U^{-1}\left|\psi_{\rm rot}\right>,
\label{left_right}
\ee
from which we obtain
\be
i\hbar\frac{\p\left|\psi_{\rm rot}\right>}{\p t}=\left(U\cH U^{-1}-i\hbar U\frac{\p U^{-1}}{\p t}\right)
\left|\psi_{\rm rot}\right>\equiv \cH_{\rm rot}\left|\psi_{\rm rot}\right>
\label{standard_Schroedinger}
\ee
by multiplying $U$ to Eq. (\ref{left_right}) from the left. From Eq. (\ref{standard_Schroedinger}) 
we can immediately read off the Hamiltonian in the rotating frame:
\be
\cH_{\rm rot}=U\cH U\D-i\hbar U\frac{\p U\D}{\p t}=U\cH U\D+i\hbar\frac{\p U}{\p t}U\D.
\ee
If we insert $U(t)=e^{-i\omega tI_z}$, we get
\be
\cH_{\rm rot}=-(g_N\mu_NH_z-\hbar\omega)I_z+g_N\mu_NH_\bot I_x.
\ee
which is completely time-independent. In the next section we will show
that under the special condition of quadratic anisotropy we can transform a Hamiltonian
with more than one oscillating term into a Hamiltonian that is
also completely time-independent.

\begin{figure}[htb]
  \begin{center}
    \leavevmode
\epsfxsize=8cm
\epsffile{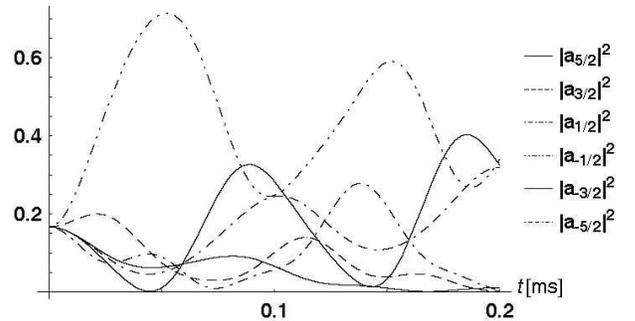}
  \end{center}
\caption{Grover algorithm calculated by means of Eq.~(\protect\ref{general_rotating_frame_approx}) 
in the QC scheme, where  
$\left|s\right>=(1/\protect\sqrt{6})\sum_{m=-5/2}^{5/2}\left|m\right>$ is concentrated mainly
into $\left|-1/2\right>$ after 0.05 ms for $h_1=h_2=h_3=h_4=h_5=\hbar\delta_3/2g_N\mu_N=5$ G,
$\delta_1=\delta_2=\delta_4=\delta_5=0$. }
\label{Grover_I=52_M=-12}
\end{figure}

\section{The generalized rotating frame}
\label{sec_generalized}

For the case where the Hamiltonian $\cH$ contains many magnetic fields $H_{x,k}(t)$
that oscillate at different frequencies $\omega_k$,
the transformation to the standard rotating frame cannot be used anymore.
In this section we develop a concept that allows us to 
transform the Hamiltonian $\cH$ into a generalized rotating frame, i.e.
$\cH\rightarrow\cH_{\rm grot}$,
where $\cH_{\rm grot}$ is completely time-independent.
Then the time evolution of the nuclear spins can be calculated non-perturbatively
in this generalized rotating frame.
In contrast to previous
work\cite{Leuenberger&Loss} our method also holds for vanishing
detuning energies $\hbar\delta_k\rightarrow 0$, which is essential to perform 
non-perturbative unitary operations.  
Once the control over $2I$ magnetic fields is established, the scheme
proposed here allows for quantum information processing and quantum storage with a {\em single}
pulse, provided that there is sufficient signal
amplification due to the spin ensemble.
The only requirements are that the quadrupolar splitting $A$ is much
larger than the detuning energies $\delta_k$, so that only
the most left diagram in Fig.~\ref{nuclear_spin_scheme}
governs the time evolution of the nuclear spins (see Sec.~\ref{sec_perturbative}).

\begin{figure}[htb]
  \begin{center}
    \leavevmode
\epsfxsize=8cm
\epsffile{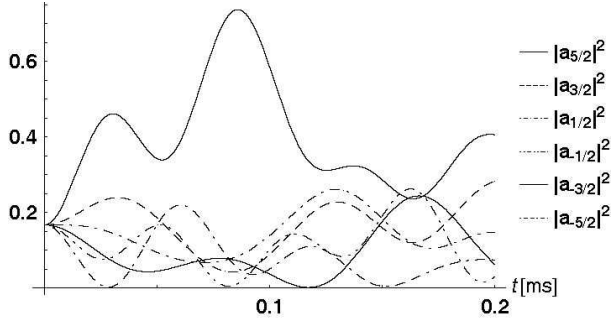}
  \end{center}
\caption{Grover algorithm calculated by means of Eq.~(\protect\ref{general_rotating_frame_approx}) 
in the QC scheme, where  
$\left|s\right>=(1/\protect\sqrt{6})\sum_{m=-5/2}^{5/2}\left|m\right>$ is concentrated mainly
into $\left|-3/2\right>$ after 0.05 ms for $h_1=h_2=h_3=h_4=h_5=\hbar\delta_4/2g_N\mu_N=5$ G,
$\delta_1=\delta_2=\delta_3=\delta_5=0$. }
\label{Grover_I=52_M=-32}
\end{figure}

We first start with the example for two detuning energies $\hbar\delta_1$
and $\hbar\delta_2$ and give the general derivation for arbitrary
spin lengths $I$ afterwards. The corresponding Hamiltonian for three states reads
\be
\cH^{(2)}=\left[\begin{array}{ccc} 
\varepsilon_a & \left<a\left|V\right|b\right> & 0  \\
\left<b\left|V\right|a\right> & \varepsilon_b & \left<b\left|V\right|c\right> \\
0 & \left<c\left|V\right|b\right> & \varepsilon_c \ea,
\label{3DHamiltonian}
\ee
where $\left|a\right>$, $\left|b\right>$, and $\left|c\right>$ are three consecutive
spin states ($a,b,c$ are integers or half-integers with $a=b+1=c+2$, $-I\le a,b,c\le I$)
with their eigenenergies $\varepsilon_a,\varepsilon_b,\varepsilon_c$.
For example $a=3/2$, $b=1/2$, and $c=-1/2$, which is shown in Fig. \ref{nuclear_spin_scheme}.
If the quadratic anisotropy $A$ is much larger than the detuning energies $\hbar\delta_k$,
we can neglect all the transition diagrams with blue detuning energies shown in 
Fig. \ref{nuclear_spin_scheme}. Applying also the rotating wave approximation leaves us
with
\be
\cH_{\rm app}^{(2)}=\left[\begin{array}{ccc} 
\varepsilon_a & h_1e^{i\omega_1t+i\Phi_1} & 0  \\
h_1e^{-i\omega_1t-i\Phi_1} & \varepsilon_b & h_2e^{i\omega_2t+i\Phi_2} \\
0 & h_2e^{-i\omega_2t-i\Phi_2} & \varepsilon_c \ea,
\label{3DHamiltonian_approx}
\ee
where $h_1=H_1\left<a\left|I_x\right|b\right>$, $h_2=H_2\left<b\left|I_x\right|c\right>$.

The goal is now to find a unitary transformation $U$ that renders
our Hamiltonian $\cH_{\rm app}^{(2)}$ time-independent.
We know from the previous section that the transformed Hamiltonian
has the form
\be
\cH_{\rm grot}=U\cH_{\rm app} U\D-i\hbar U\frac{\p U\D}{\p t}=U\cH_{\rm app} U\D+i\hbar\frac{\p U}{\p t}U\D.
\label{3DHamiltonian_grot}
\ee
Instead of inserting $U(t)=e^{-i\omega tI_z}$,
we make an ansatz with three parameters $\omega_a$, $\omega_b$, and
$\omega_c$:
\be
U=\left[\begin{array}{ccc} 
e^{-i\omega_at} & 0 & 0  \\
0 & e^{-i\omega_bt} & 0 \\
0 & 0 & e^{-i\omega_ct} \ea.
\ee

\begin{figure}[htb]
  \begin{center}
    \leavevmode
\epsfxsize=8cm
\epsffile{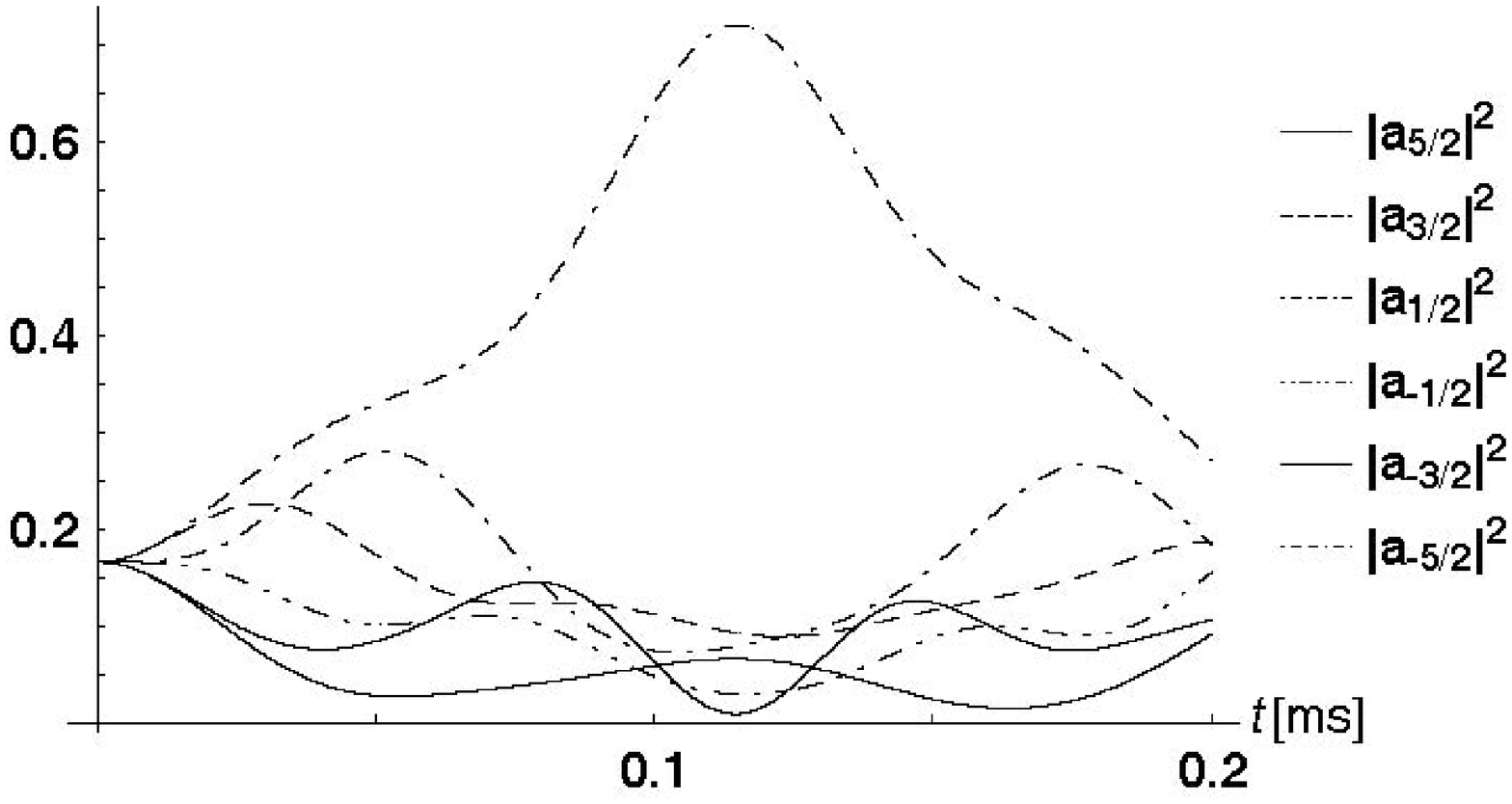}
  \end{center}
\caption{Grover algorithm calculated by means of Eq.~(\protect\ref{general_rotating_frame_approx}) 
in the QC scheme, where  
$\left|s\right>=(1/\protect\sqrt{6})\sum_{m=-5/2}^{5/2}\left|m\right>$ is concentrated mainly
into $\left|-5/2\right>$ after 0.05 ms for $h_1=h_2=h_3=h_4=h_5=\hbar\delta_5/2g_N\mu_N=5$ G,
$\delta_1=\delta_2=\delta_3=\delta_4=0$. }
\label{Grover_I=52_M=-52}
\end{figure}

Evaluation of $U\cH_{\rm app}^{(2)}U\D$ yields
\bea
\cH_{\rm grot}^{(2)} & = & \left[\begin{array}{ccc} 
\varepsilon_a & h_1e^{i\Omega_1t} & 0  \\
h_1e^{-i\Omega_1t} & \varepsilon_b & h_2e^{i\Omega_2t} \\
0 & h_2e^{-i\Omega_2t} & \varepsilon_c \ea \nn\\
& & +i\hbar\frac{\p U}{\p t}U\D,
\label{3DHamiltonian_approx2}
\eea
with the abbreviations $\Omega_1=-\omega_a+\omega_b+\omega_1+\Phi_1/t$ 
and $\Omega_2=-\omega_b+\omega_c+\omega_2+\Phi_2/t$.
Since we want to make the Hamiltonian in Eq. (\ref{3DHamiltonian_approx2})
time-independent, we require that 
\bea
\Omega_1 & = & -\omega_a+\omega_b+\omega_1+\Phi_1/t=0, \label{eq1} \\
\Omega_2 & = & -\omega_b+\omega_c+\omega_2+\Phi_2/t=0. \label{eq2}
\eea
In this way we get also rid of the Phases $\Phi_1$ and $\Phi_2$.
Solving the Eqs. (\ref{eq1}) and (\ref{eq2})
gives
\bea
\omega_a & = & \omega_1+\omega_2+\omega_c+(\Phi_1+\Phi_2)/t, \label{eq1_solved} \\
\omega_b & = & \omega_2+\omega_c+\Phi_2/t. \label{eq2_solved}
\eea
$\omega_c$ is a free parameter, which we choose to be 
$\omega_c=-(\omega_1+\omega_2)/2-(\Phi_1+\Phi_2)/2t$.
Then Eqs. (\ref{eq1_solved}) and (\ref{eq2_solved}) simplify to
\bea
\omega_a & = & (\omega_1+\omega_2)/2+(\Phi_1+\Phi_2)/2t, \\
\omega_b & = & (\omega_2-\omega_1)/2+(\Phi_2-\Phi_1)/2t.
\eea
Then the second term in Eq. (\ref{3DHamiltonian_grot}) is
\be
i\hbar\frac{\p U}{\p t}U\D=\left[\begin{array}{ccc} 
\frac{\hbar(\omega_1+\omega_2)}{2} & 0 & 0  \\
0 & \frac{\hbar(\omega_2-\omega_1)}{2} & 0 \\
0 & 0 & -\frac{\hbar(\omega_1+\omega_2)}{2} \ea.
\ee

\begin{figure}[htb]
  \begin{center}
    \leavevmode
\epsfxsize=8cm
\epsffile{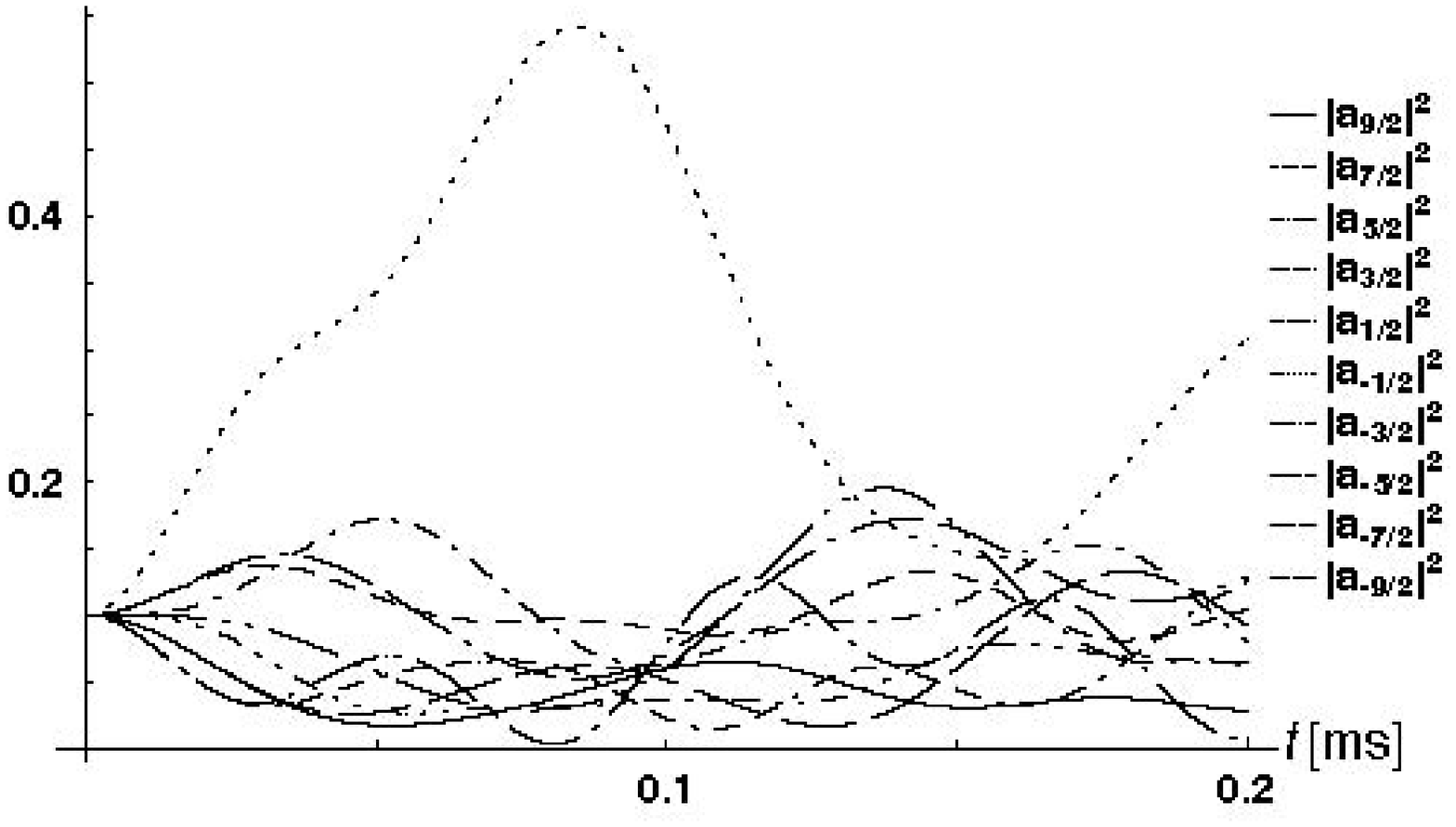}
  \end{center}
\caption{Grover algorithm calculated by means of Eq.~(\protect\ref{general_rotating_frame_approx}) 
in the QC scheme, where  
$\left|s\right>=(1/\protect\sqrt{10})\sum_{m=-9/2}^{9/2}\left|m\right>$ is concentrated mainly
into $\left|-1/2\right>$ after 0.05 ms for $h_1=h_2=h_3=h_4=h_5=h_6=h_7=h_8=h_9=\hbar\delta_5/2g_N\mu_N$ $=5$ G,
$\delta_1=\delta_2=\delta_3=\delta_4=\delta_6=\delta_7=\delta_8=\delta_9=0$. }
\label{Grover_I=92_M=-12}
\end{figure}

If we insert the external frequencies 
$\omega_1=\varepsilon_b/\hbar-\varepsilon_a/\hbar-\delta_1$
and $\omega_2=\varepsilon_c/\hbar-\varepsilon_b/\hbar+\delta_1-\delta_2$, we obtain
\be
\cH_{\rm grot}^{(2)}=\left[\begin{array}{ccc} 
\frac{\varepsilon_a+\varepsilon_c-\hbar\delta_2}{2} & h_1 & 0  \\
h_1 & \frac{\varepsilon_a+\varepsilon_c+\hbar(2\delta_1-\delta_2)}{2} & h_2 \\
0 & h_2 & \frac{\varepsilon_a+\varepsilon_c+\hbar\delta_2}{2} \ea,
\label{3DHamiltonian_grot_prefinal}
\ee
which is equivalent to
\be
\cH_{\rm grot}^{(2)}=\left[\begin{array}{ccc} 
0 & h_1 & 0  \\
h_1 & \hbar\delta_1 & h_2 \\
0 & h_2 & \hbar\delta_2 \ea,
\label{3DHamiltonian_grot_final}
\ee
where we have subtracted $\frac{\varepsilon_a+\varepsilon_c-\hbar\delta_2}{2}{\rm Id}$.
Id is the identity matrix.

Now we give the general derivation for the transformation
to the generalized rotating frame for arbitrary spin lengths $I$.
From the above derivation, we know that for a spin of arbitrary length $I$
the Hamiltonian in the generalized rotating frame has the form
\bea
\cH_{\rm grot}^{(2I)} & = & \left[\begin{array}{ccccc}
\ddots & \ddots &  &  &  \\
\ddots & \varepsilon_{m_5} & h_5e^{i\Omega_5t} &  & \\
 & h_5e^{-i\Omega_5t} & \varepsilon_{m_6} & h_6e^{i\Omega_6t} &  \\
 &  & h_6e^{-i\Omega_6t} & \varepsilon_{m_7} & \ddots \\
 &  &  & \ddots & \ddots \ea \nn\\
& & +i\hbar\frac{\p U}{\p t}U\D,
\label{2IDHamiltonian_approx}
\eea
which has non-zero entries only in the diagonal and first off-diagonal
lines. $\left|m_0\right>=\left|2I\right>,\left|m_1\right>=\left|2I-1\right>,
\ldots,\left|m_{2I}\right>=\left|-2I\right>$ are the eigenstates
of the Hamiltonian shown in Eq.~(\ref{Hamiltonian_0}) with eigenenergies
$\varepsilon_{m_0},\varepsilon_{m_1},\ldots,\varepsilon_{m_{2I}}$.
The unitary transformation reads
\be
U=\left[\begin{array}{ccccc} 
\ddots & & & & \\
& e^{-i\omega_{m_5}t} &  &  & \\
& & e^{-i\omega_{m_6}t} & & \\
& &  & e^{-i\omega_{m_7}t} & \\
& & & & \ddots \ea,
\ee
with only diagonal elements that are non-zero.

\begin{figure}[htb]
  \begin{center}
    \leavevmode
\epsfxsize=8cm
\epsffile{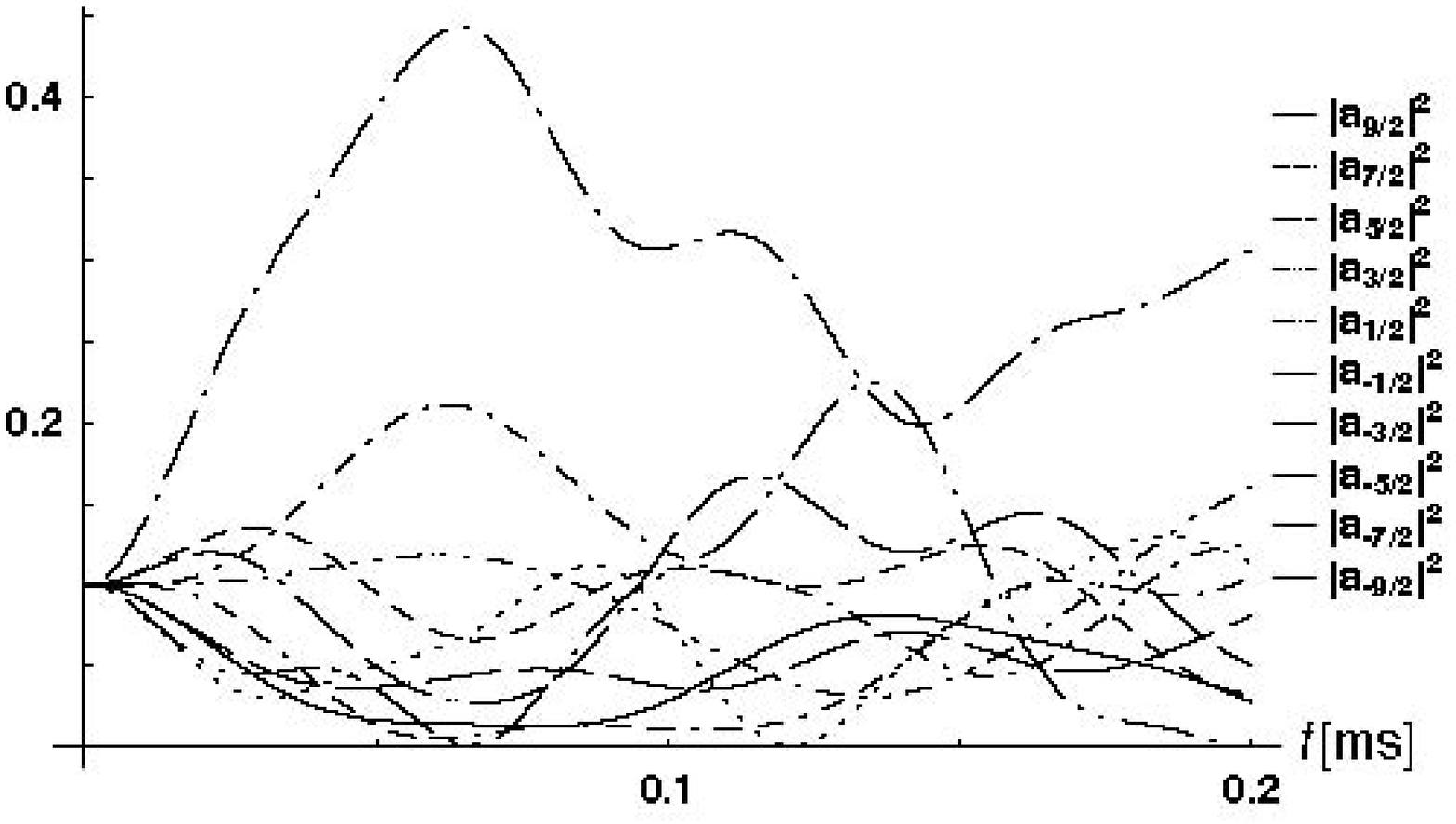}
  \end{center}
\caption{Grover algorithm calculated by means of Eq.~(\protect\ref{general_rotating_frame_approx}) 
in the QC scheme, where  
$\left|s\right>=(1/\protect\sqrt{10})\sum_{m=-9/2}^{9/2}\left|m\right>$ is concentrated mainly
into $\left|-3/2\right>$ after 0.05 ms for $h_1=h_2=h_3=h_4=h_5=h_6=h_7=h_8=h_9=\hbar\delta_6/2g_N\mu_N$ $=5$ G,
$\delta_1=\delta_2=\delta_3=\delta_4=\delta_5=\delta_7=\delta_8=\delta_9=0$. }
\label{Grover_I=92_M=-32}
\end{figure}

For $\cH_{\rm grot}^{(2I)}$ to be time-independent we have to
solve $2I$ linear equations of the form
\bea
\Omega_1 & = & -\omega_{m_0}+\omega_{m_1}+\omega_1+\Phi_1/t=0, \label{lineq1} \\
\Omega_2 & = & -\omega_{m_1}+\omega_{m_2}+\omega_2+\Phi_2/t=0, \label{lineq2} \\
& \vdots & \nn\\
\Omega_5 & = & -\omega_{m_5}+\omega_{m_6}+\omega_5+\Phi_5/t=0, \label{lineq5} \\
\Omega_6 & = & -\omega_{m_6}+\omega_{m_7}+\omega_6+\Phi_6/t=0, \label{lineq6} \\
& \vdots & \nn\\
\Omega_{2I-1} & = & -\omega_{m_{2I-2}}+\omega_{m_{2I-1}}+\omega_{2I-1}+\Phi_{2I-1}/t=0, \label{lineq2I-1} \\
\Omega_{2I} & = & -\omega_{m_{2I-1}}+\omega_{m_{2I}}+\omega_{2I}+\Phi_{2I}/t=0, \label{lineq2I}
\eea
Summing over all linear equations (\ref{lineq1}) to (\ref{lineq2I}) leads to
\be
\omega_{m_0}=\omega_{m_{2I}}+\sum\limits_{k=1}^{2I}(\omega_k+\Phi_k/t),
\ee
where $\omega_{m_{2I}}$ can be chosen arbitarily.
Inserting this result back into the linear equations (\ref{lineq1}) to (\ref{lineq2I}) gives
\bea
\omega_{m_1} & = & \omega_{m_{2I}}+\sum\limits_{k=2}^{2I}(\omega_k+\Phi_k/t), \\
\omega_{m_2} & = & \omega_{m_{2I}}+\sum\limits_{k=3}^{2I}(\omega_k+\Phi_k/t), \\
& \vdots & \nn\\
\omega_{m_5} & = & \omega_{m_{2I}}+\sum\limits_{k=6}^{2I}(\omega_k+\Phi_k/t), \\
\omega_{m_6} & = & \omega_{m_{2I}}+\sum\limits_{k=7}^{2I}(\omega_k+\Phi_k/t), \\
& \vdots & \nn\\
\omega_{m_{2I-1}} & = & \omega_{m_{2I}}+\sum\limits_{k=2I}^{2I}(\omega_k+\Phi_k/t).
\eea

\begin{figure}[htb]
  \begin{center}
    \leavevmode
\epsfxsize=8cm
\epsffile{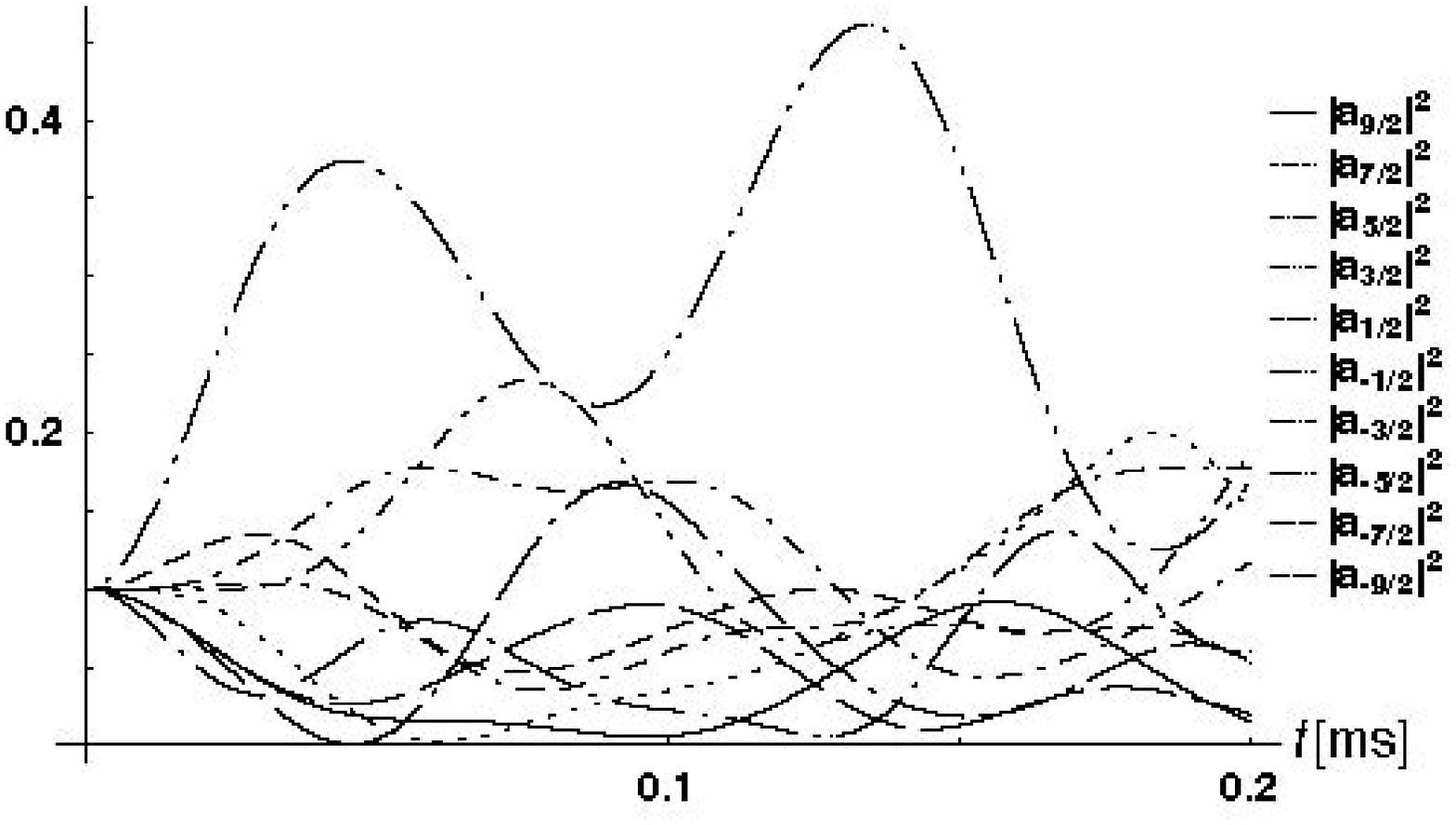}
  \end{center}
\caption{Grover algorithm calculated by means of Eq.~(\protect\ref{general_rotating_frame_approx}) 
in the QC scheme, where  
$\left|s\right>=(1/\protect\sqrt{10})\sum_{m=-9/2}^{9/2}\left|m\right>$ is concentrated mainly
into $\left|-5/2\right>$ after 0.05 ms for $h_1=h_2=h_3=h_4=h_5=h_6=h_7=h_8=h_9=\hbar\delta_7/2g_N\mu_N=5$ G,
$\delta_1=\delta_2=\delta_3=\delta_4=\delta_5=\delta_6=\delta_8=\delta_9=0$. }
\label{Grover_I=92_M=-52}
\end{figure}

Then the second term in Eq. (\ref{2IDHamiltonian_approx}) $i\hbar\frac{\p U}{\p t}U\D$
has the diagonal elements $\hbar(\omega_{m_{2I}}+\sum_{k=1}^{2I}\omega_k),
\hbar(\omega_{m_{2I}}+\sum_{k=2}^{2I}\omega_k),\ldots,
\hbar(\omega_{m_{2I}}+\sum_{k=2I}^{2I}\omega_k),\hbar\omega_{m_{2I}}.$
All the off-diagonal elements of $i\hbar\frac{\p U}{\p t}U\D$ are zero.
The external frequencies are
\bea
\omega_1 & = & \varepsilon_{m_1}/\hbar-\varepsilon_{m_0}/\hbar-\delta_1 \\
\omega_2 & = & \varepsilon_{m_2}/\hbar-\varepsilon_{m_1}/\hbar-\delta_2+\delta_1 \\
& \vdots & \nn\\
\omega_5 & = & \varepsilon_{m_5}/\hbar-\varepsilon_{m_4}/\hbar-\delta_5+\delta_4 \\
\omega_6 & = & \varepsilon_{m_6}/\hbar-\varepsilon_{m_5}/\hbar-\delta_6+\delta_5 \\
& \vdots & \nn\\
\omega_{2I} & = & \varepsilon_{m_{2I}}/\hbar-\varepsilon_{m_{2I-1}}/\hbar-\delta_{2I}+\delta_{2I-1}
\eea
After setting $\omega_{m_{2I}}=0$, we then obtain for the diagonal elements of 
$i\hbar\frac{\p U}{\p t}U\D$ the following results:
$(-\varepsilon_{m_0}+\varepsilon_{m_{2I}}-\hbar\delta_{2I}),
(-\varepsilon_{m_1}+\varepsilon_{m_{2I}}+\hbar\delta_1-\hbar\delta_{2I}),
(-\varepsilon_{m_2}+\varepsilon_{m_{2I}}+\hbar\delta_2-\hbar\delta_{2I}),\ldots,
(-\varepsilon_{m_{2I-1}}+\varepsilon_{m_{2I}}+\hbar\delta_{2I-1}-\hbar\delta_{2I}),0.$
Finally, subtracting $(\varepsilon_{m_{2I}}-\hbar\delta_{2I})$Id from $\cH_{\rm grot}^{(2I)}$
in Eq. (\ref{2IDHamiltonian_approx}) yields
the time-independent Hamiltonian
\be
\cH_{\rm grot}^{(2I)}=\left[\begin{array}{ccccc} 
0 & h_1 & 0 & \cdots & 0 \\
h_1 & \hbar\delta_1 & h_2 & \ddots & \vdots \\
0 & h_2 & \hbar\delta_2 & \ddots & 0  \\
\vdots & \ddots & \ddots & \ddots & h_{2I} \\ 
0 & \cdots & 0 & h_{2I} & \hbar\delta_{2I} \ea.
\label{general_rotating_frame_approx}
\ee
This Hamiltonian allows us to evaluate the time evolution of the nuclear spin
system for arbitrary spin lengths $I$ and arbitrary times $t$ non-perturbatively.
Note that the Hamiltonian in Eq.~(\ref{general_rotating_frame_approx}) remains valid
even in the limit $\delta_k\rightarrow 0$, where the perturbation
expansions for multiphoton (more than 1) transition amplitudes, such as in Eqs.~(\ref{S3}) and (\ref{S2}), break down.
However, we must require that $|g_N\mu_N H_k|\ll |A|$, which means that
the larger $|A|$, the faster the quantum information processing (see Ref.~\onlinecite{Leuenberger2002}).
The time evolution of the state $\left|\psi(t)\right>$ reads
\be
\left|\psi(t)\right>=U^\dagger(t)e^{-i\cH_{\rm grot}^{(2I)}t/\hbar}\left|\psi(t=0)\right>.
\ee 
Propagators of the
form $U^\dagger(t)e^{-i\cH_{\rm grot}^{(2I)}t/\hbar}$ have $2I$
phases $\Phi_k$ and $2I$ detunings
$\hbar\delta_k$, which determine the $2I$
phases and the $2I$ moduli of $a_m$.
We subtracted two
degrees of freedom: the global phase and the normalization condition,
respectively.
In this way we can produce a state $\left|\psi(t)\right>$
with arbitrary amplitudes $a_m$.

\begin{figure}[htb]
  \begin{center}
    \leavevmode
\epsfxsize=8cm
\epsffile{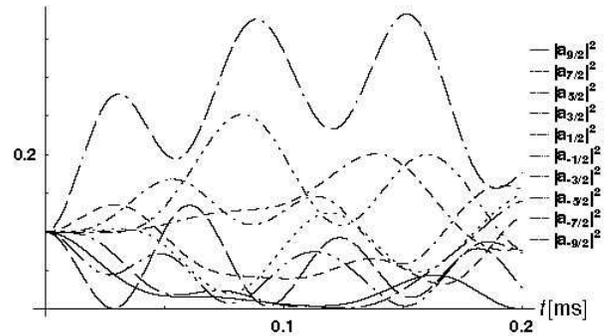}
  \end{center}
\caption{Grover algorithm calculated by means of Eq.~(\protect\ref{general_rotating_frame_approx}) 
in the QC scheme, where  
$\left|s\right>=(1/\protect\sqrt{10})\sum_{m=-9/2}^{9/2}\left|m\right>$ is concentrated mainly
into $\left|-7/2\right>$ after 0.05 ms for $h_1=h_2=h_3=h_4=h_5=h_6=h_7=h_8=h_9=\hbar\delta_8/2g_N\mu_N=5$ G,
$\delta_1=\delta_2=\delta_3=\delta_4=\delta_5=\delta_6=\delta_7=\delta_9=0$. }
\label{Grover_I=92_M=-72}
\end{figure}

\section{The Grover algorithm}
\label{sec_Grover}

Now we make use of the non-perturbative calculation method shown
in the previous section to compute the Grover algorithm.
We start from a configuration where mainly the ground state
$\left|3/2\right>$ is populated, see Fig.~\ref{nuclear_spin_scheme}.
This can be achieved by the Overhauser effect\cite{Overhauser}. 
Then we produce an equal superposition of the nuclear spin states
$\left|s\right>=(1/\sqrt{n})\sum_m\left|m\right>$, which represents
the initial state of the Grover algorithm.
$n$ is the number of basis states involved in the search.
For example we can prepare the state $\left|s\right>=(1/\sqrt{3})\sum_{m=-1/2}^{3/2}\left|m\right>$
with the parameters $H_1=H_2=1$ G,
$H_3=0$, $\delta_1=6083$ s$^{-1}$, and $\delta_2=0$,
which is shown in Fig.~\ref{QC}.
In Fig.~\ref{QC_num} we calculated the time evolution of 
$\left|s\right>$ exactly by solving the Schr\"odinger equation
numerically. This calculation reveals small oscillations
that are mainly due to the five neglected diagrams in Fig.~\ref{nuclear_spin_scheme}.

\begin{figure}[htb]
  \begin{center}
    \leavevmode
\epsfxsize=8cm
\epsffile{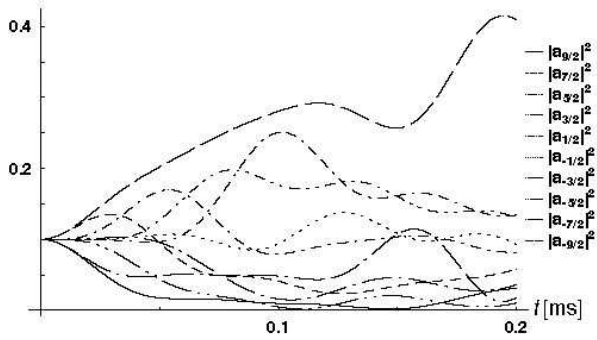}
  \end{center}
\caption{Grover algorithm calculated by means of Eq.~(\protect\ref{general_rotating_frame_approx}) 
in the QC scheme, where  
$\left|s\right>=(1/\protect\sqrt{10})\sum_{m=-9/2}^{9/2}\left|m\right>$ is concentrated mainly
into $\left|-9/2\right>$ after 0.2 ms for $h_1=h_2=h_3=h_4=h_5=h_6=h_7=h_8=h_9=\hbar\delta_9/2g_N\mu_N=5$ G,
$\delta_1=\delta_2=\delta_3=\delta_4=\delta_5=\delta_6=\delta_7=\delta_8=0$. }
\label{Grover_I=92_M=-92}
\end{figure}

Next, we choose one basis state $\left|M\right>$ to be the one
that we are looking for. Since we encode the information into
the eigenenergies $\delta_{I-m}$ of the eigenstates $\left|m\right>$
in the generalized rotating frame, we can choose $\delta_{I-m}=0$
for all $m$, except for $M$, i.e. $\delta_{I-M}\ne 0$.
In order to find $\left|M\right>$, we make it degenerate with the
state $\left|s\right>$ in the generalized rotating frame. 
So $\delta_{I-M}=\left<s\left|\cH_{\rm grot}^{(2I)}\right|s\right>\ne 0$.
As we want to obtain the highest speed for the quantum information processing,
the best choice for the Zeeman fields is to make them all equal, i.e. 
$h_1=h_2=\ldots=h_{2I}=:h$.
The degeneracy condition now leads to
\be
\left<s\left|\cH_{\rm grot}^{(2I)}\right|s\right>
=\frac{1}{n}(2n-2)h+\frac{1}{n}\sum_{k=1}^{2I}\hbar\delta_k
\stackrel{!}{=}\hbar\delta_{I-M},
\ee
from which we obtain directly
\be
h=\frac{\hbar\delta_{I-M}}{2}.
\ee
We continue the above example, where we have prepared the state 
$\left|s\right>=(1/\sqrt{3})\sum_{m=-1/2}^{3/2}\left|m\right>$.
After making $\left|s\right>$ degenerate with $\left|-1/2\right>$
in the generalized rotating frame,
i.e. $\left<s\left|\cH_{\rm grot}^{(3)}\right|s\right>=\hbar\delta_2$,
the amplitude concentrates mainly into the state $\left|-1/2\right>$
after 0.2 ms.
In contrast to Refs.~\onlinecite{Farhi,Grover&Sengupta},
the Hamiltonian in Eq.~(\ref{general_rotating_frame_approx}) has only nearest-neighbor couplings,
which results in a decreasing amplification of $\left|M\right>$
with increasing $I$ or $|M|$. However, even for the largest nuclear spin
$I=9/2$, we find that the resolution for identifying $\left|M\right>$
is still sufficient, i.e. greater than 10\% (see below).
Again, we have also calculated this time evolution exactly by solving 
the Schr\"odinger equation numerically, which is shown in Fig.~\ref{Grover_num}.
The small oscillations due to the five neglected diagrams in Fig.~\ref{nuclear_spin_scheme}
are now almost invisible. Thus the larger the quadrupolar splitting $A$,
the better the agreement between our non-perturbative method and
exact numerics.

\begin{figure}[htb]
  \begin{center}
    \leavevmode
\epsfxsize=5cm
\epsffile{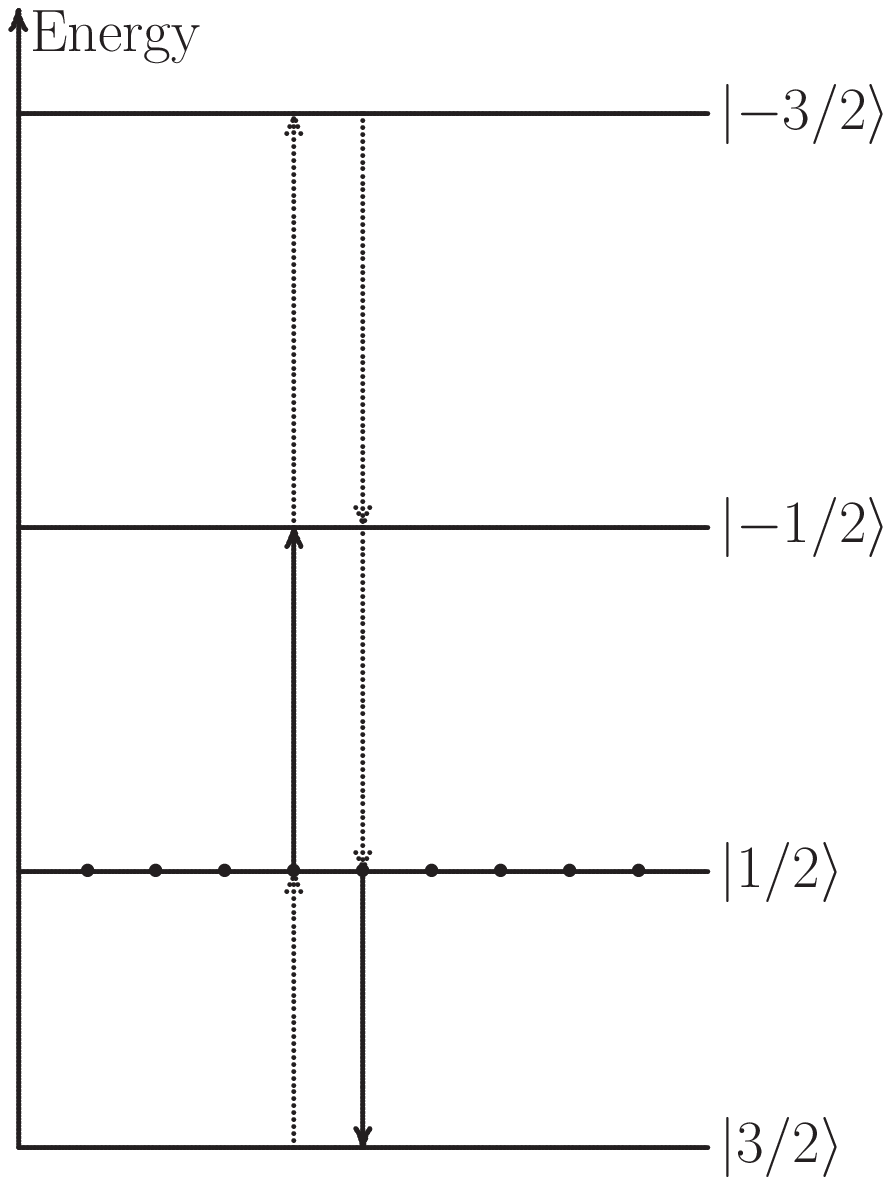}
  \end{center}
\caption{In this example only the state $\left|1/2\right>$
is fully populated after performing the Grover algorithm.
For the readout of this state, we have to irradiate the nuclear spin system
with magnetic fields that oscillate at frequencies that match exactly
the level separations.}
\label{nuclear_spin_readout}
\end{figure}

We have also calculated the Grover sequence for spin lengths up to $I=9/2$.
In all the cases for all spin lengths up to $I=9/2$ and for 
all the searchable states $-I\le M\le I$ the Grover sequence
leads to an amplification of the amplitude of $\left|M\right>$
between about 10\% and 100\%.
In Figs.~\ref{Grover_I=32_M=12} and \ref{Grover_I=32_M=-32}
the spin has length $I=3/2$ and the searched state
is $\left|M=1/2\right>$ and $\left|M=-3/2\right>$, respectively.
When we increase the spin length $I$,
the most interesting elements for applications in semiconductors
are $^{27}$Al, $^{55}$Mn, and $^{67}$Zn with nuclear spin $I=5/2$,
and $^{73}$Ge and $^{113}$In with nuclear spin $I=9/2$.
Therefore we have computed the Grover sequence
for $I=5/2$ and $M=-1/2,-3/2,-5/2$, which can be seen
in Figs.~\ref{Grover_I=52_M=-12}, \ref{Grover_I=52_M=-32}, \ref{Grover_I=52_M=-52},
respectively,
and also for $I=9/2$ and $M=-1/2,-3/2,-5/2,-7/2,-9/2$, which is shown
in Figs.~\ref{Grover_I=92_M=-12}, \ref{Grover_I=92_M=-32}, \ref{Grover_I=92_M=-52},
\ref{Grover_I=92_M=-72}, \ref{Grover_I=92_M=-92},
respectively.
Because of the symmetry of the Hamiltonian in Eq.~(\ref{general_rotating_frame_approx})
along the diagonal, the two computations for $\pm M$ look always the same.
Hence, we have shown the Grover sequence only
for half of the states $-I\le M<0$.

\section{Readout of the result}
\label{sec_readout}

After we have performed the Grover quantum search algorithm, we wish to extract the
final state that we have been searching for.
This can be achieved by using conventional pulsed NMR, where no coherence is required.
Fig.~\ref{nuclear_spin_readout} demonstrates an example where only the
state $\left|1/2\right>$ is completely populated.
If we irradiate the sample with magnetic fields $H_{x,k}(t)$
with zero detuning energies $\delta_k=0$ $\forall k$,
where the external frequencies are
tuned in such a way that they match the energy level spaces, i.e.
$\omega_{I-m}=\varepsilon_{m+1}-\varepsilon_m$,
then we induce only emission from $\left|1/2\right>$ to $\left|3/2\right>$
and absorption from $\left|1/2\right>$ to $\left|-1/2\right>$.
The emission and absorption spectrum shown in Fig.~\ref{nuclear_spin_absorption}
identifies unambigously the state $\left|1/2\right>$.
In this way we can read out any information stored in the nuclear spin system.

\begin{figure}[htb]
  \begin{center}
    \leavevmode
\epsfxsize=7cm
\epsffile{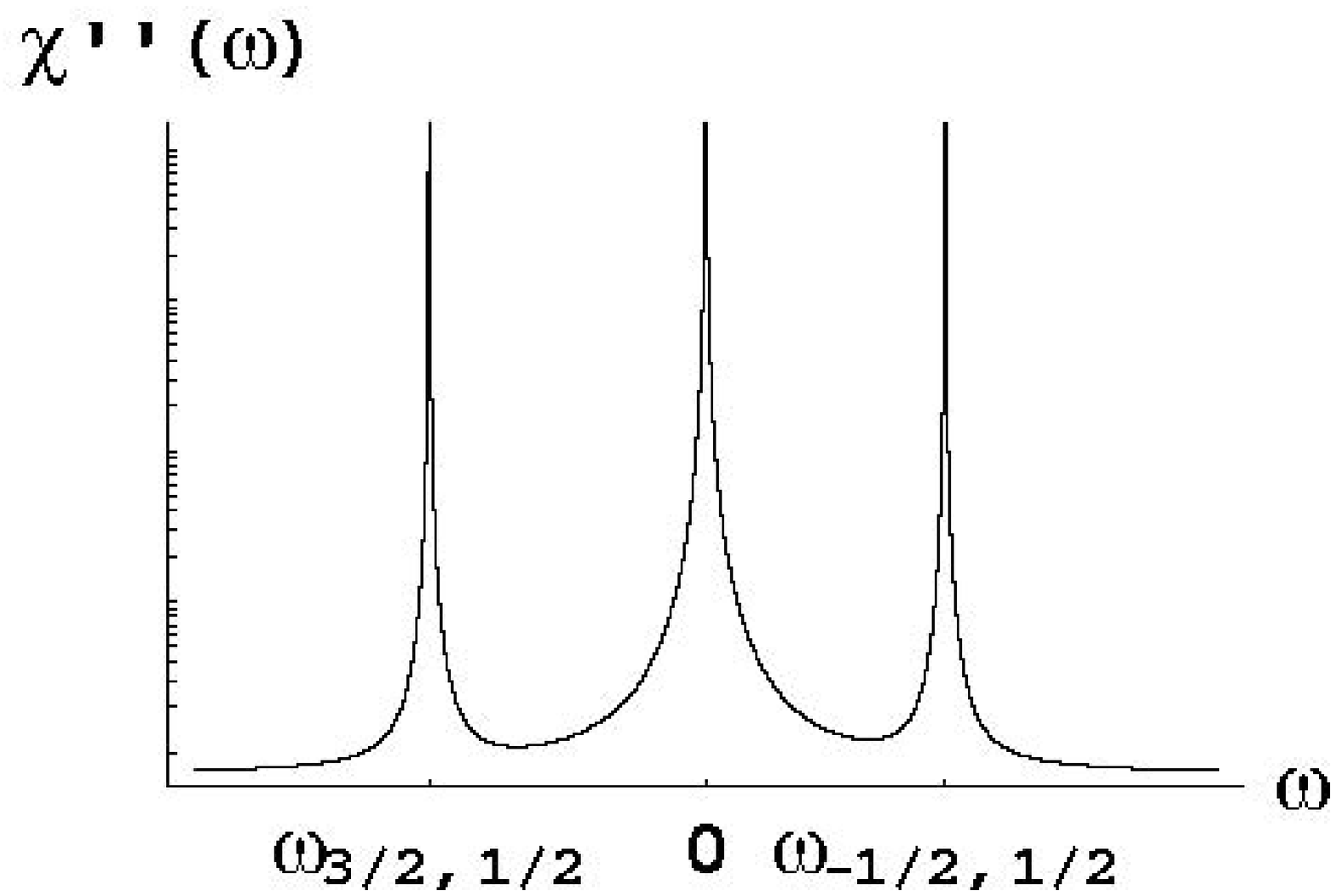}
  \end{center}
\caption{
The absorption and emission spectrum identifies then unambigously
the state $\left|1/2\right>$ for the example shown in Fig.~\protect\ref{nuclear_spin_readout}.}
\label{nuclear_spin_absorption}
\end{figure}

\section{Conclusion}

In our theoretical proposal we have shown that the Grover algorithm
can be performed in the nuclear spin system of semiconductors.
The main requirement for the Grover algorithm to work is the anisotropy in the spin system,
which is provided by the quadrupolar splitting.
The reason is that the quadrupolar splitting makes the energy levels non-equidistant,
which renders the nuclear spin states distinguishable.
Only then we can find an effective Hamiltonian of the form shown in
Eq.~(\ref{general_rotating_frame_approx}) that describes the coherent time
evolution of the nuclear spin ensemble for arbitrary spin lengths and times.
It turned out that the larger the quadrupolar splitting,
the better the agreement between our non-perturbative method using
the effective Hamiltonian and exact numerical calculations.
So the larger the symmetry breaking of the energy spacings,
the better the control over the coherent time evolution of
the nuclear spin system.

The first test for the experimental feasibility of our proposal
would be the implementation of multiphoton Rabi oscillations,
as proposed in Ref.~\onlinecite{Leuenberger2002}.
This can be achieved by applying only one oscillating magnetic field.
Fig.~\ref{nutation} shows a 2-photon Rabi oscillation between 
the states $\left|3/2\right>$ and $\left|-1/2\right>$
for a nuclear spin $I=3/2$.
In general, multiphoton Rabi oscillations can be thought of
as nutation of the large spin between the spin states $\left|m\right>$.

Once our scheme works experimentally, it would be interesting
to apply a longitudinal magnetic field that has a large gradient.
In this case the semiconductor sample could be divided into
several regions, each of which could be addressed separately
by different transverse magnetic fields.
In this way one could perform parallel quantum information
processes in each of the regions.
Instead of using a large gradient field,
it would also be possible to vary the g-factor in the sample,
as was shown in the experiment of Ref.~\onlinecite{Salis2001} for the electron spin
in Al$_x$Ga$_{1-x}$As, where the variation of the concentration
of Al $x$ leads to a variation of the electron g-factor.

\begin{figure}[htb]
  \begin{center}
    \leavevmode
\epsfxsize=7cm
\epsffile{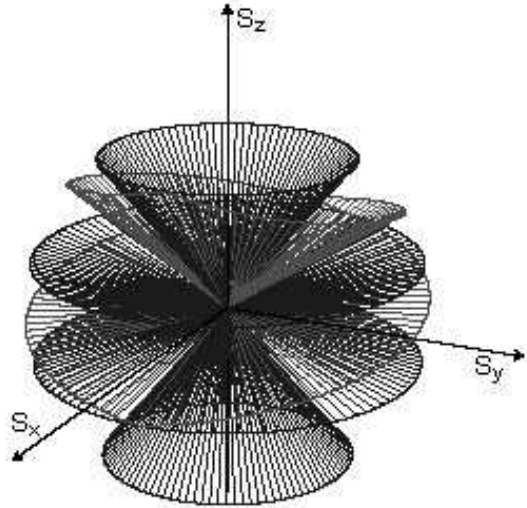}
  \end{center}
\caption{
2-photon Rabi oscillation between the states $\left|3/2\right>$ and $\left|-1/2\right>$
for a nuclear spin $I=3/2$.}
\label{nutation}
\end{figure}

\section{Acknowledgement} 

We thank M. E. Flatt\'e for useful comments.
We acknowledge the Swiss NSF, NCCR Nanoscience, and the US
NSF and DARPA for financial support.

%end PRB

\end{document}